\documentstyle[aps,amssymb,twocolumn]{revtex}

\input{tcilatex}

\begin{document}
\title{Electrostatic Tuning of the Superconductor-Insulator Transition in Two
Dimensions}
\author{Kevin A. Parendo, K. H. Sarwa B. Tan, A. Bhattacharya, M. Eblen-Zayas, N. E.
Staley, and A. M. Goldman}
\address{School of Physics and Astronomy, University of Minnesota, 116 Church St. SE,%
\\
Minneapolis, MN 55455, USA}
\date{
\today%
}
\maketitle

\begin{abstract}
Superconductivity has been induced in insulating ultra-thin films of
amorphous bismuth using the electric field effect. The screening of the
electron-electron interaction was found to increase with electron
concentration in a manner correlated with the tendency towards
superconductivity. This does not preclude an increase in the density of
states being important in the development of superconductivity. The
superconductor-insulator transition appears to belong to the universality
class of the three dimensional XY model.
\end{abstract}

\pacs{PACS numbers: ()}

Superconductor-insulator (SI) transitions in ultrathin films\cite{Goldman}
are believed to be examples of quantum phase transitions (QPTs)\cite{Sachdev}%
. Such transitions occur at zero temperature in response to the tuning of an
external parameter of a system that alters its ground state. Although many
systems exhibit QPTs, including $^{4}$He adsorbed on random substrates,
two-dimensional electron gases, and numerous complex strongly correlated
electron materials including high temperature superconductors, the SI
transition is one of the most fundamental because of its apparent connection
with the phase-number uncertainty relation of quantum mechanics\cite
{Phillips}. The standard approaches to tuning SI transitions involve
increasing film thickness and applying perpendicular magnetic fields. Both
of these can introduce complications. Increasing film thickness changes
carrier density but may also alter the disorder landscape. Applying
perpendicular magnetic fields introduces vortices and the complexity of
vortex physics.

Here we report the electrostatic tuning of the SI transition in ultra-thin
amorphous bismuth ({\it a-}Bi) films using the electric field effect. This
tuning changes the carrier density without altering the disorder landscape.
Modulation of superconductivity using the field effect is an old idea \cite
{HebardFiory}, which has been revived by recent work on high temperature
(cuprate) superconductors\cite{Ahn,Cassin}. However, these studies have not
revealed significant features of the transitions between the various ground
states. The difficulties of preparing high quality cuprate films with
thicknesses on the order of the electrostatic screening length and
chemically doped so that they are near a phase boundary has limited the
usefulness of these studies. These constraints are either irrelevant or
easily satisfied in work on homogeneously disordered ultra-thin films.

A single-crystal of (100) SrTiO$_{3}$ (STO) served as both the substrate and
the gate insulator. Its unpolished surface was mechanically thinned \cite
{Bhattacharya}, leaving parallel surfaces approximately 50 $\mu $m apart.
Platinum was pre-deposited at 300 K forming a 1000 \AA\ thick gate electrode
on the thinned back surface and 100 \AA\ thick measurement electrodes on the
epi-polished front surface. The substrate was then placed in a dilution
refrigerator/UHV deposition apparatus \cite{Hernandez}. A 10 \AA\ thick
underlayer of amorphous antimony ({\it a-}Sb) followed by subsequent layers
of {\it a-}Bi were deposited {\em in situ} under UHV conditions ($\thicksim
10^{-9}$ torr) through shadow masks. The substrates were held at
liquid-helium temperatures during deposition. Films grown in this manner are
believed to be homogeneously disordered \cite{Strongin}.

Resistances were measured by employing standard four-probe DC methods using
a 1 nA current, a value well within the linear regime of the I-V
characteristics. A voltage source provided the gate-film bias. All
electrical leads were filtered at 300 K with RC filters to attenuate 60 Hz
noise and $\pi $-section filters to attenuate RF noise, and at the mixing
chamber of the refrigerator with 2 m long Thermocoax\ cables \cite{Zorin} to
attenuate GHz noise. The measurement and deposition processes were
alternated, with the film always kept below 14 K in a UHV environment,
permitting the study of the evolution of electrical transport properties
with thickness.

In the sequence of films in which electrostatic gating was studied, the film
thickness was varied from 9.60 \AA\ to 10.22 \AA\ in four steps. All films
were insulating, with $R(T)$ exhibiting a temperature dependence consistent
with Mott variable range hopping. All curves exhibited resistance flattening
for $T\lesssim 60$ mK. We will argue later that this behavior is not
evidence of an intermediate metallic regime. It is believed to be a
consequence of the electrons in the film failing to cool despite significant
efforts to shield from external noise and thermal radiation.

The 10.22 \AA\ thick film was the first in the sequence in which an
electrostatically tuned SI transition (Fig. 1) could be induced. Resistance
decreased with increasing positive gate voltage ($V_{G}$) at all
temperatures, consistent with previous observations that the carriers in
metallic {\it a-}Bi are electrons \cite{Buckel}. A logarithmic dependence on
temperature became a better fit at about 7 V and continued to do so for all
larger gate voltages at high temperatures despite the presence of
superconducting behavior at low temperatures.

One can qualitatively distinguish between insulating and superconducting
ground states by examining the sign of $dR/dT$ at low temperatures. At low V$%
_{G}$, $dR/dT<0$, suggesting an insulating state. With increasing $V_{G}$, $%
dR/dT$ eventually changed sign, suggesting a superconducting state. Zero
resistance, within the limits of the scatter of the data, was observed for V$%
_{G}$ 
\mbox{$>$}%
\ 38 V. The highest temperature at which zero resistance was observed
increased monotonically with $V_{G}$, from 52 mK at 38V to 59 mK at 42.5 V.
At 42.5 V, all effects due to gating saturated, so R(T) for $V_{G}>$ $42.5$
V was identical to R(T) at 42.5V.

The response of the resistance to $V_{G}$ was stronger at low temperatures
than at high temperatures, implying that the high temperature conductance
was affected much less by electrostatic charging than the superconducting
pairing mechanism. The major changes in $R(T)$ in response to $V_{G}$
appeared only as low temperature deviations from the weakly insulating $%
V_{G}=0$ curve. Qualitatively similar behavior has been found in the
perpendicular magnetic field-tuned SI transition of In$_{2}$O$_{3}$ films 
\cite{Samband}, as well as in studies of the metal-insulator transition \cite
{Abrahams}.

The inset of Fig. 1 shows the field effect for both signs of gate voltage at
65 mK for the 10.22 \AA\ thick film. Negative $V_{G}$ produces a small
change in film resistance, while positive $V_{G}$ causes a large decrease.
At higher temperatures, all R(T) curves for negative $V_{G}$ largely
coincide with the $V_{G}=0$ curve. Thus hopping transport is only changed a
small amount. Similarly, for a 10.69 \AA\ thick film (not shown) with a 446
mK transition temperature, the asymmetry in the response for the two signs
of gate voltage was evident in that for $V_{G}=50$ V, $T_{c}$ increased to
502 mK, while for $V_{G}=-50$ V, it decreased only to 436 mK. This film was
prepared in an earlier sequence of depositions in which the
superconductor-insulator transition was overshot. 

Measurements at low temperatures using thinned STO as a gate dielectric have
demonstrated that electron transfer is approximately linear with gate
voltage from zero at $V_{G}=0$ to approximately $3\times 10^{13}$ cm$^{-2}$
at $V_{G}=$ $50$ V \cite{Bhattacharya} and it does not saturate above that
value. Thus, the observed saturation of film behavior cannot result from
properties of the gate dielectric. The film's intrinsic electron
concentration is unknown, as the geometry in this particular set of studies
was limited to parallel magnetic fields, precluding Hall effect measurements.

The simplicity of electrostatic charging as a tuning parameter provides an
opportunity to analyze data in ways not possible for either the thickness or
magnetic field tuned SI transitions. The behavior of R(T) at high
temperatures, as a function of gate voltage, can reveal features of the
disordered 2D electron system relevant to its superconductivity. In 2D
metals, in zero magnetic field, the conductance is the sum of the classical
Boltzmann conductance ($G_{B}$) and corrections due to coherent
backscattering/weak localization ($G_{WL}$) \cite{Ramak} and
electron-electron interactions ($G_{EE}$) \cite{Altschuler}. Both produce a
decrease in conductance that goes as the logarithm of temperature. This is
based on a perturbative analysis valid when $k_{F}l$ $\gg $ 1, where $k_{F}$
is the 2D Fermi wavevector and $l$ is the electronic mean free path.
However, it is possible that this analysis may work even when $k_{F}l\gtrsim
1$. (The value of $k_{F}l$ for these films, estimated from their high
temperature conductivity is approximately 1.6.) Thus, 
\[
G_{WL}(T)+G_{EE}(T)=[\alpha p+(1-\frac{3}{4}F^{*})]\frac{e^{2}}{2\pi ^{2} 
\rlap{\protect\rule[1.1ex]{.325em}{.1ex}}h%
}ln(T) 
\]
The weak localization correction is determined by the temperature exponent
of the inelastic scattering time, p, and the localization parameter, $\alpha 
$. The interaction term is controlled by a screening parameter, $F^{*}$,
which increases from 0 to 0.866 as the electron-electron interaction changes
from being unscreened to completely screened. The inset to Fig. 2 shows a
plot of $G$ vs. $ln(T)$ for the $V_{G}=$ 12.5 V curve of the 10.22 \AA\
thick film. This curve is nonlinear at low T because the system in tending
towards superconductivity. For T 
\mbox{$>$}%
175 mK, the curve is linear in $ln(T)$, and the coefficient of the logarithm
depends on the values of $\alpha p$ and $F^{*}$. Assuming the weak
localization contribution, $\alpha p,$ to be constant for all $V_{G}$, by
subtracting slopes of $G(ln(T))$ at nonzero $V_{G}$ from the slope of $%
G(ln(T))$ at $V_{G}=0,$ one can determine the value of $\Delta
F^{*}=F^{*}(V_{G})-F^{*}(V_{G}=0)$. This function, which is then the
increase in screening constant induced by gate voltage, relative to the
screening constant intrinsic to the ungated film, is shown in figure 2. $%
F^{*}$ increases monotonically by approximately 0.35 from $V_{G}=0$ to $%
V_{G}=42.5$ V. $F^{*}$ will increase with this functional form, no matter
the precise value of $\alpha p$ or $F^{*}(V_{G}=0)$. (Note that in this
analysis, we force a logarithmic fit to the exponential insulating curves.
This is fairly accurate at temperatures higher than roughly 150 mK.) A
positive magnetoresistance, logarithmic in B, that would allow determination
of $\alpha p$ and of $F^{*}(V_{G}=0)$ is expected in fields $B>>k_{B}T/g\mu
_{B}$. We observed positive magnetoresistance for all B at 300 mK at $%
V_{G}=10$ V, but no specific field dependence could be determined.

The increase in $F^{*}$ with $V_{G}$ is correlated with the enhancement of
the tendency towards superconductivity. In conventional metals the latter is
controlled by the competition between the Coulomb repulsion and the
phonon-mediated attractive interaction. Increased screening leads to the
suppression of Coulomb repulsion, ultimately permitting superconductivity to
develop. Once superconductivity is achieved, increased screening should move 
$T_{c}$ to higher temperatures. Saturation of both $T_{c}$ and $F^{*}$ with $%
V_{G}$ might not be unexpected in the limit of strong metallicity, as the
density of states for a 2D metal is a constant independent of carrier
concentration. The electron density determines the value of conductivity at
high temperatures. Since a major change in the low temperature behavior
occurs without a substantial change in the resistance at higher
temperatures, screening, which changes significantly, and not the areal
charge density, might be controlling the transition to superconductivity.
However, in the BCS model, the product of the density of states and the
interaction potential determines the transition temperature. Therefore, an
increase in the density of states may also contribute to the inducing of
superconductivity.

The asymmetry of the response to the sign of $V_{G}$ would suggest that the
Fermi energy of the ungated film is close to the mobility edge. Increasing
the electron concentration with positive $V_{G}$ would then increase the
screening and the density of states significantly, whereas decreasing the
electron concentration with negative $V_{G}$\ would result in small changes
in the hopping transport, screening and the density of states.

It is expected that conductivity data associated with continuous QPTs can be
collapsed using a finite-size scaling analysis \cite{Sachdev}. As data has
been acquired at increasingly lower temperatures, such analyses have been
observed to fail because of flattening of the curves of R(T). Scaling also
appears to fail when curves of R(T) that exhibit the onset of
superconducting behavior fan down from weakly insulating curves. In the
present work, finite size scaling of R(T) data with $V_{G}$ as the tuning
parameter was successful because we confined the temperature range and used
fine increments of values of the control parameter. All data for $T\leq 60$
mK was excluded, assuming that the observed flattening of R(T) was due to a
failure to cool the electrons. In the inset of Fig. 3, $R$($V_{G}$) is
plotted at fixed temperature for fifteen isotherms between 65 mK and 100 mK.
This yields a well-defined crossing point at $V_{G}=11$ V, which is taken to
be the critical value of the tuning parameter, $V_{GC}$ . The critical
resistance ($R_{C}$) of 19,100 $\Omega $ is large in comparison with values
found for the SI transition with other tuning parameters. Plotting $R/R_{C}$
versus $|V_{G}-V_{GC}|T^{-1/\nu z}$ for 54 values of $V_{G}$ between 0 and
42.5 V, the product $\nu z=2/3$ is found to yield the best collapse of data
onto two separate branches, one for superconducting films and the other for
insulating films. This is shown in Fig. 3. Assuming the value of $z$ to be
1, the value $\nu =2/3$ is consistent with the universality classes of
either the three dimensional (3D) XY or inverse XY models. The latter is
believed to describe the transition to superconductivity in 3D \cite
{Kiometzis} and might be expected for a QPT in a system governed by the 2D
XY model at nonzero temperature with $z=1$ \cite{Sachdev}. This value of the
exponent product is different from those found using thickness as a tuning
parameter for {\it a-}Bi films \cite{Goldman} and magnetic field as a tuning
parameter for InO$_{x}$ \cite{Hebard} and MoGe \cite{Yazdani} films, but is
consistent with a study of the magnetic field tuned transition in a-Bi films 
\cite{Markovic}.

The breakdown of scaling when data above 0.1 K is included suggests that the
critical regime only extends from zero to some low temperature on the order
of 0.1 K. Except for the limited temperature range, the quality of the curve
is as good as or better than that of earlier scaling analyses, from the
highest value of R/R$_{C}$ down to 0.6 and for a range of $%
|V_{G}-V_{GC}|T^{-1/vz}$ extending from 0 to over 700. The analysis is shown
down to $R/R_{C}=0$ to demonstrate how isotherms deviate from the
superconducting scaling branch. The bottom curve is for the lowest
temperature and appears to adhere to the scaling function. Curves generated
from data taken at increased temperatures deviate at higher values of $%
R/R_{C}$. This implies that scaling works best at the lowest temperatures.
In this work, all R(T)\ curves tend toward becoming temperature independent
below approximately 60 mK. Curves for which 11 V $<\ V_{G}<$ 38 V have
nonzero resistance at the lowest measured temperature, while curves for
which $V_{G}$ $\geq $ 38 V show zero resistance. The fact that all R(T)
curves from both groups satisfy the same scaling function for the same $\nu
z $ suggests that these two regimes are not different. This is the main
reason for our earlier assertion that {\em in this instance} the observed
flattening of resistance is not intrinsic. In other experimental situations,
the flattening could be a real metallic phase.

Any scaling analysis has a lower temperature bound. Thus, data that appear
on the insulating branch may actually belong on the superconducting one, as
R(T) curves that appear to be insulating might actually be observed to be
superconducting if lower temperatures became accessible. The continuous
nature of the scaling function at the zero of $|V_{G}-V_{GC}|$ would allow
for a higher critical resistance than that identified here without change of
the critical exponent product.

The electrostrictive response of STO can be ruled out as an explanation of
the observed effects because $R(V_{G})$ is asymmetric in $V_{G}$ while the
electrostrictive response of STO is symmetric \cite{Bhattacharya}.

The thinness (10\AA ) of the {\it a-}Sb layer rules out the possibility that
screening may be caused by an accumulation of a mobile charge at the
interface between the STO substrate and the {\it a-}Sb layer, similar to the
field-effect modulation found between STO and Al$_{2}$O$_{3}$ \cite{Ueno}.
The fact that negative gate voltage causes little if any change in the film
is different from what was found in previous work on a-Bi grown on {\it a-}
Ge \cite{Markovic1}.

In summary, the electric field effect has been used to tune the 2D SI
transition. The evolution of superconductivity is correlated with an
increase in screening. The size of the critical regime appears to be quite
small. The exponent product $\nu z\thicksim 2/3$ is found, so the
universality class of the transition appears to be that of the 3D XY model.
This product is different from the value $\nu z\thicksim 4/3$ that has been
found in numerous previous experiments, which has been suggested to be a
signature of classical percolation in 2D \cite{Kapitulnik}. It is possible
that the underlying mechanism for this electrostatically tuned transition is
different from those in studies that are tuned by changing film thickness or
applying magnetic fields.

The authors would like to acknowledge very useful conversations with Boris
Shklovskii, Anatoly Larkin, Leonid Glazman, Oriol Valls, and Alex Kamenev.
This work was supported in part by the National Science Foundation under
grant NSF/DMR-0138209.


\begin{figure}[tbp]
\caption[Evolution of$R(T)$ of the 9.19\AA\ film as a function of in-plane
magnetic field. Field values from top to bottom are: 12.5, 12, 11.6, 11.5,
11, 10, 9, 8, 7, 6, 5, 4, 3, 2, and 0 T. Inset: temperature at which dR/dT
becomes zero is plotted vs. applied field.]{Change in the screening
parameter $\Delta F^{*}$ versus $V_{G}$. Inset: Conductance versus $\ln (T)$
for the $V_{G}=12.5$ V curve.}
\label{fig2}
\end{figure}
\begin{figure}[tbp]
\caption{Scaling analysis of $R(T)$ with $V_{G}$ as tuning parameter. The
critical exponent product $\nu z=2/3$. Inset: $R(V_{G})$ for isotherms from
65 to 100 mK, yielding a distinct crossing point at the critical gate
voltage of $11$ V.}
\label{fig3}
\end{figure}

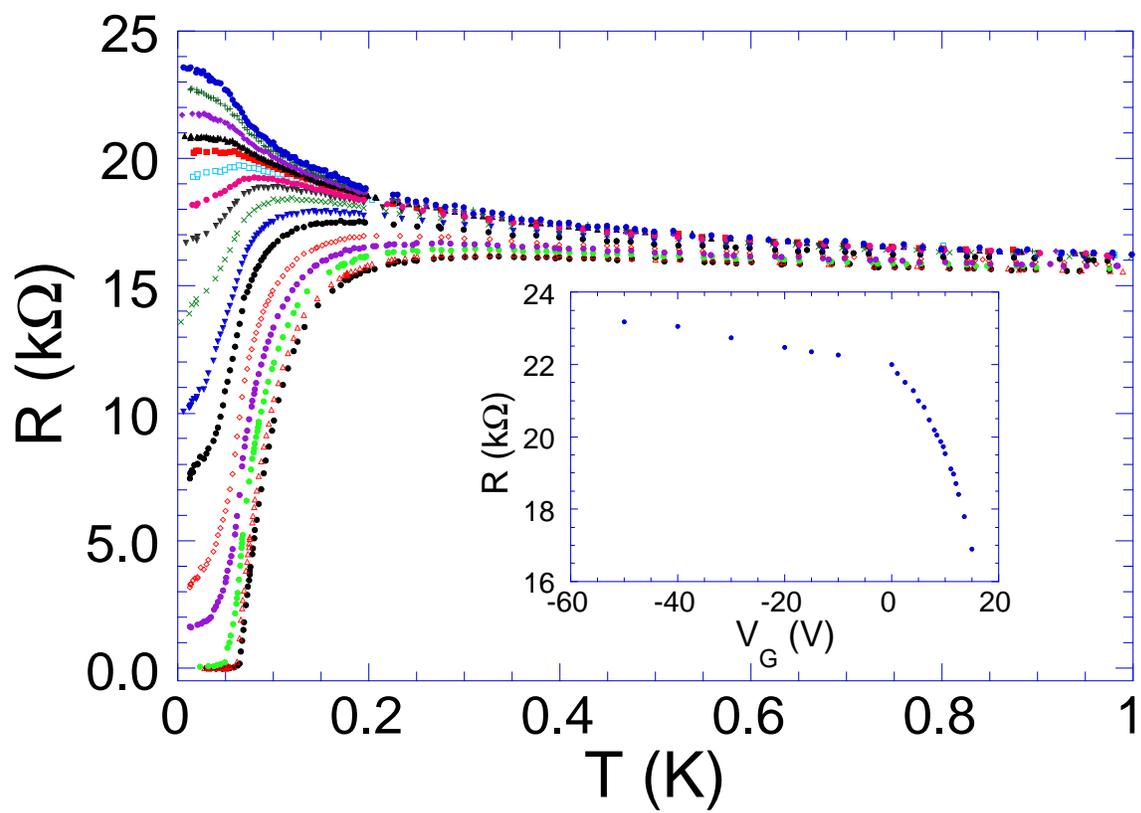
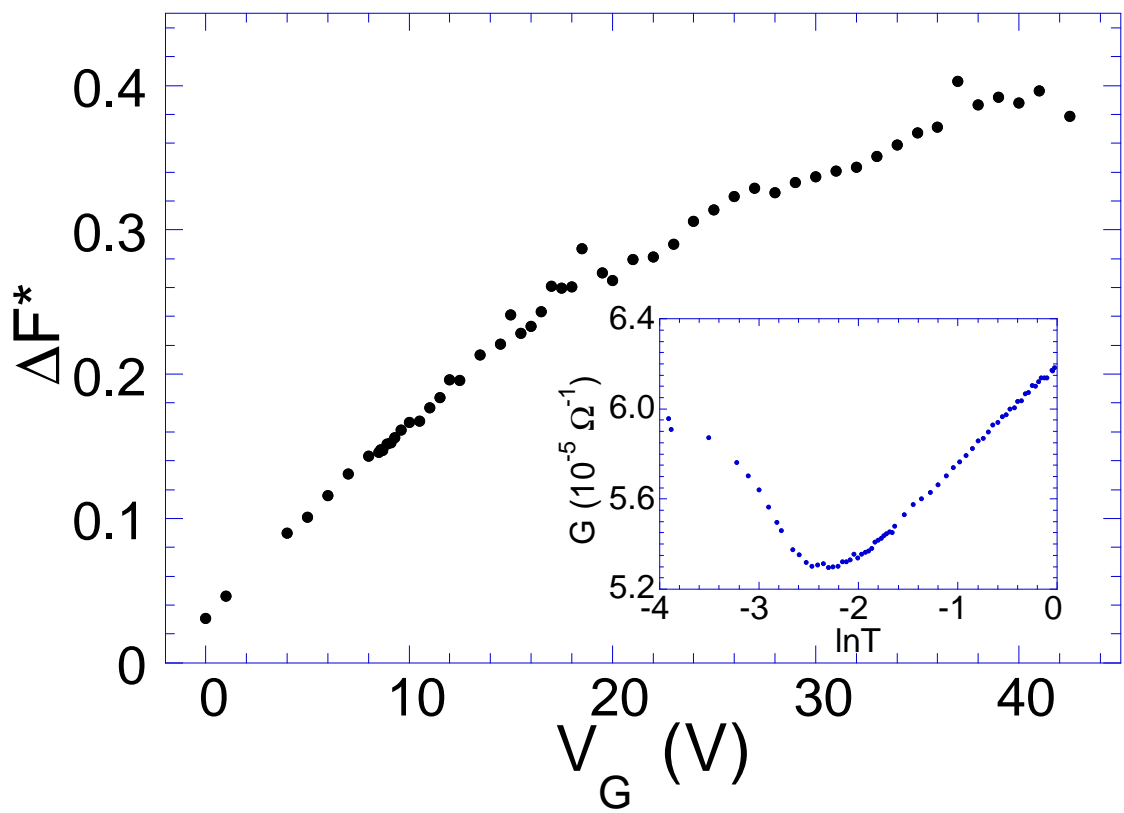
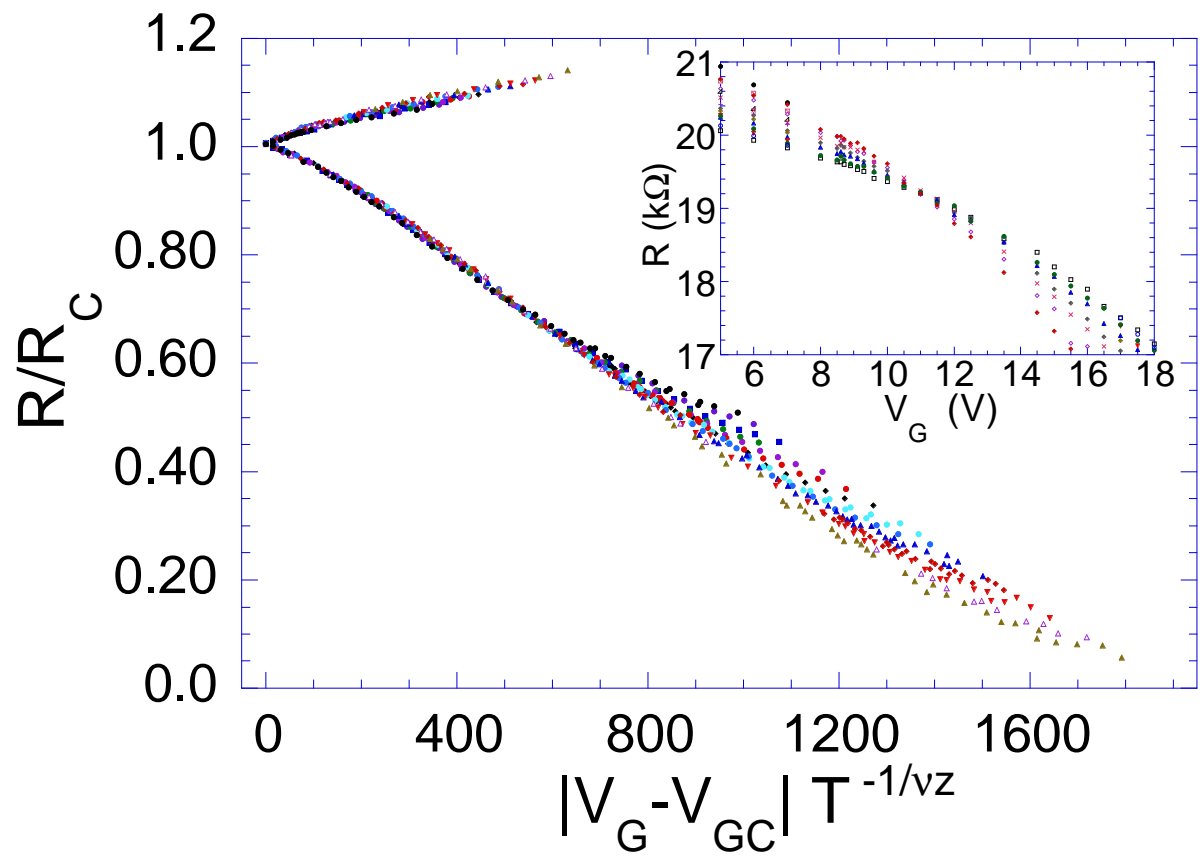
\end{document}